\documentstyle[epsf,aps,prb,multicol]{revtex}

\begin{document}
\title{Analytic static structure factors and pair--correlation
functions \\
 for the unpolarized homogeneous electron gas}

\author{Paola Gori-Giorgi, Francesco Sacchetti}
\address{Dipartimento di Fisica and Unit\`a INFM, Universit\`a di Perugia,
         Via A. Pascoli 1, 06123 Perugia, Italy}
\author{Giovanni B. Bachelet}
\address{Dipartimento di Fisica and Unit\`a INFM,
         Universit\`a di Roma ``La Sapienza'', Piazzale Aldo Moro 2, 00185
         Rome, Italy}

\date{\today}

\maketitle

\begin{abstract}
We propose a simple and accurate model for the electron
static structure factors (and corresponding pair--correlation
functions) of the 3D unpolarized homogeneous electron gas.
Our spin--resolved pair--correlation function is
built up with a combination of analytic constraints and fitting
procedures to quantum Monte Carlo data, and,  in comparison to
previous attempts (i) fulfills more known integral and differential
properties of the exact pair--correlation function, (ii) is analytic
both in real and in reciprocal space, and (iii) accurately
interpolates the newest, extensive diffusion--Monte
Carlo data of Ortiz, Harris and Ballone [Phys. Rev. Lett.
{\bf 82}, 5317 (1999)]. This can be of interest for
the study of electron correlations of real materials and for the construction
of new exchange and correlation energy density functionals.
\end{abstract}
\begin{multicols}{2}


\section{Introduction}

The homogeneous electron gas, a model solid whose positive ionic
charges are smeared throughout the whole crystal volume to yield a
shapeless, uniform positive background (whence the nickname of
jellium) has provided, since the very start of quantum mechanics, a
key conceptual reference and a mine of information for solid--state
and many--body theorists.\cite{wigner,ymarch,histo}  Initially it was
mostly regarded as an approximation of the true distribution of
valence electrons in simple metals, since, in spite of its crudity, it
could already account for some of their experimental
properties.\cite{kitzim}  Although the importance of valence--charge
inhomogeneites in real materials was soon recognized (and described
first by perturbation\cite{pert} and later by self--consistent
pseudopotential theory\cite{hsc-bhs}), the homogeneous electron gas
stood by itself, over the decades, as an independent active field of
theoretical\cite{review} and
numerical\cite{cepald,Pickett,OB,cepmorsen,kwon,OBH}
investigation.  One reason for this continued interest is that the
model, by ignoring the ionic lattice which makes real materials
different from one another, allows the theorists to concentrate on key
aspects of the electron--electron interaction.  Another reason for
caring about such an unrealistic system resides in its connection to
the inhomogeneous electron gas:\cite{wigner,ymarch,slater}
not only does the jellium model
represent an obvious limit, but also, through the Density Functional
Theory \cite{hks} and its Local Density Approximation, it links to a
popular and very successful description of real materials.\cite{nobel}
For the latter reason, from the simplest Hartree--Fock
approximation \cite{hks} to the pioneering QMC
simulations,\cite{cepald} almost any theory of jellium, its electron
correlations and its pair--correlation functions has also implied an
improved understanding and construction of Kohn--Sham energy
functionals.\cite{gunlun,PZ}

In this context our work aims at a new simple analytic expression for the
pair--correlation function of the homogeneous electron gas, which
describes the spatial correlations of electron pairs with prescribed
spin orientations.  A good model pair--correlation function and
static structure factor has its own
interest; its availability over a wide density range is crucial
for new developments and applications of the Density Functional
Theory, through the construction of {\em ab initio} exchange
and correlation energy functionals in generalized gradient
approximations\cite{GGA} and in other beyond--LDA 
schemes.\cite{gunlun,WDA,Chacon,Dobson}
As a consequence, over the last 20 years, several
authors already proposed ingenious expressions for this or related
functions.\cite{Chacon,Kimball2,Ichimaru,Sacchetti,Lee,Becke,Valone,PerdewWang,Gritsenko,Proynov}
A first motivation for resuming and
improving over previous efforts is the avalaibility, from recent
quantum Monte Carlo (QMC) simulations,\cite{OB,OBH} 
of a wealth of new numerical results for the
pair--correlation functions and static structure factors of jellium.
A second motivation comes from
the observation that most of the previous models
were not spin--resolved and all of them neither fulfilled all
the known exact properties, nor
were given in analytic, closed form both in real and
reciprocal space.  Our goal is thus to give a new, spin--resolved
expression for the pair--correlation function which is analytic both
in real and reciprocal space, automatically incorporates more
exact properties than any previous expression, and contains
enough free parameters to fit the new QMC results.\cite{OBH}

We recall the exact properties of the pair--correlation
function for the unpolarized jellium in Sec.~\ref{sec_properties}.
The three subsequent sections  are devoted to
a description of our general strategy (Sec.~\ref{sec_strat})
and of the resulting functional form for the antiparallel-- (Sec.~\ref{sec_ud})
and parallel--spin (Sec.~\ref{sec_uu}) pair--correlation functions.
In Sec.~\ref{sec_fitQMC} we describe our fitting procedure to
QMC data.\cite{OBH}
Once the exact constraints are imposed, 18 free parameters (9 for
antiparallel spins and 9 for parallel spins)
are enough to yield extremely accurate two--dimensional
fits of the $\approx 9000+9000$ new QMC data points\cite{OBH}
as a function of the interelectronic distance $r$
and the density parameter $r_s$ in the relevant density range
$r_s\le 10$.
Our results
are discussed and compared with the widely used Perdew--Wang\cite{PerdewWang}
model in Sec.~\ref{sec_results}.

In Sec.~\ref{sec_correnergy} we report the correlation energy
which corresponds to our model pair--correlation function, and we find that
its agreement with the QMC
energies,\cite{OBH} not targeted by our fitting procedure, is
as good ($\sim 5$\%) as the most
popular interpolation formulae for the correlation energy. 
The last Sec.~\ref{concper} is devoted to conclusions and perspectives.
\section{Exact properties}
\label{sec_properties}
We briefly recall many known properties of the pair correlation
function of the unpolarized homogeneous electron gas.
Its integral properties
(sum rules) will be rewritten in terms of $q \to 0$ properties of
Fourier transforms, since this choice turns out to be convenient for our
subsequent steps. Hartree atomic units are used throughout this work.

\subsection{Definitions}
For an electronic system the pair--correlation function $g_{\sigma_1
\sigma_2}({\bf r}_1,{\bf r}_2)$, if $n_{\sigma}({\bf r})$ is the
density of electrons with spin $\sigma=\uparrow$ or $\downarrow$, is
defined by
\begin{eqnarray}
n_{\sigma_1}({\bf r}_1)n_{\sigma_2}({\bf r}_2) & &
g_{\sigma_1 \sigma_2}({\bf r}_1,{\bf r}_2)= \nonumber \\
& & \langle\Phi|\psi_{\sigma_1}^{\dagger}({\bf r}_1)
\psi_{\sigma_2}^{\dagger}({\bf r}_2)
\psi_{\sigma_2}({\bf r}_2)\psi_{\sigma_1}({\bf r}_1)|\Phi\rangle
\label{g_def}
\end{eqnarray}
and is thus related to the probability of finding two electrons of
prescribed spin orientations at positions ${\bf r}_1$ and ${\bf r}_2$.
The normalization of $g$ is such that the expected number of electrons
of spin $\sigma_2$ in the volume $dV$ at ${\bf r}_2$ when another
electron of spin $\sigma_1$ is known to be at ${\bf r}_1$ is
given by
\begin{equation}
dN({\bf r}_2\sigma_2|{\bf r}_1\sigma_1)=
n_{\sigma_2}({\bf r}_2)\,g_{\sigma_1\sigma_2}({\bf r}_1,{\bf r}_2)\,dV;
\label{eq_normg}
\end{equation}
the lack of any correlation amounts, then, to the condition
$g_{\sigma_1\sigma_2}({\bf r}_1,{\bf r}_2)=1$.
In the spin--unpolarized jellium the electronic spin density
$n_{\uparrow}({\bf r})= n_{\downarrow}({\bf r})=n/2=(8\pi
r_s^3/3)^{-1}$ is uniform in space (i.e.  independent of ${\bf r}$),
\cite{ekseptions} so $g_{\sigma_1 \sigma_2}({\bf r}_1,{\bf
r}_2)$ only depends on the distance between the two electrons $r=|{\bf
r}_1-{\bf r}_2|$.  The static structure factor $S(q)$ is directly
related to the Fourier transform of the pair--correlation function.
For an unpolarized homogeneous electron gas, after introducing the
Fermi wavevector $q_F=(3\pi^2\,n)^{1/3}=\alpha/r_s$, with $\alpha=
(9\pi/4)^{1/3}$, the scaled variables $\rho=q_F\,r$ and $k=q/q_F$ are
often convenient.  With these variables the static structure factors are
written as
\begin{eqnarray}
S_{\uparrow \downarrow}(k) & = & \frac{2}{3\,\pi}\int_0^\infty d\rho\,
\left[g_{\uparrow \downarrow}(\rho)-1\right]\,\rho^2\frac{\sin (k\,\rho)}
{k\,\rho}
\label{Sud_def} \\
S_{\uparrow \uparrow}(k) & = & 1+\frac{2}{3\,\pi}\int_0^\infty d\rho\,
\left[g_{\uparrow \uparrow}(\rho)-1\right]\,\rho^2\frac{\sin (k\,\rho)}
{k\,\rho}
\label{Suu_def}
\end{eqnarray}
and the total pair--correlation function and static structure factor
are given by:
\begin{eqnarray}
g(\rho;r_s) & = & \frac{1}{2}\left[g_{\uparrow \uparrow}(\rho;r_s)+
g_{\uparrow \downarrow}(\rho;r_s)\right] \label{eq_gtot} \\
S(k;r_s) & = & S_{\uparrow \uparrow}(k;r_s)+S_{\uparrow \downarrow}(k;r_s).
\label{eq_Stot}
\end{eqnarray}

\subsection{Pair--correlation function near $r=0$}
\label{sec_r0prop}
The behavior of $g_{\sigma_1\sigma_2}(r;r_s)$ in the $r\to 0$ limit can be
directly obtained from the many--body Schr\"odinger equation when two
electrons approach each other (cusp conditions):\cite{Kimball2,Kimball1,cusp_t}
\begin{eqnarray}
\frac{\partial}{\partial r}g_{\uparrow \downarrow}(r;r_s)\biggr|_{r \to 0}
& = &  g_{\uparrow \downarrow}(r\to 0;r_s) \ne 0
\label{cusp_ud} \\
\frac{\partial}{\partial r}g_{\uparrow \uparrow}(r;r_s)\biggr|_{r \to 0}
& = & g_{\uparrow \uparrow}(r\to 0;r_s) = 0
\label{Pauli} \\
\frac{\partial^3}{\partial r^3}g_{\uparrow \uparrow}(r;r_s)\biggr|_{r \to 0}
& = & \frac{3}{2}
\frac{\partial^2}{\partial r^2}g_{\uparrow \uparrow}(r;r_s)\biggr|_{r \to 0}
\ne 0.
\label{cusp_uu}
\end{eqnarray}
Eqs.~(\ref{cusp_ud})--(\ref{cusp_uu}) hold for any 3D--system of $N$ 
fermions interacting via the two--body repulsive Coulomb potential.

\subsection{Structure factor near $q=0$}
\label{sec_q0prop}
The conservation of particles in the system implies the relations:
\begin{equation}
S_{\uparrow \downarrow}(q\to 0;r_s)=S_{\uparrow \uparrow}(q\to 0;r_s)=0.
\label{Sq0}
\end{equation}
The asymmetry between the definitions~(\ref{Sud_def}) and (\ref{Suu_def})
leads to the two
well--known sum rules for $g_{\uparrow \downarrow}$ and $g_{\uparrow \uparrow}$
(see for instance Ref.~\CITE{Kimball2}).

The long--wavelength behavior of the total static structure factor
of Eq.~(\ref{eq_Stot}) is determined by the plasmon contribution, proportional
to $q^2$, and by
the single--pair and multipair quasiparticle--quasihole excitation
contributions, proportional to $q^5$ and $q^4$
respectively:\cite{histo,iwa}
\begin{equation}
S(q \to 0;r_s) = \frac{q^2}{2\, \omega_p(r_s)}+
C\,q^4+{\cal O}(q^5),
\label{plasmon}
\end{equation}
where $\omega_p(r_s)=\sqrt{3/r_s^3}$ is the classical plasma frequency.
In the paramagnetic gas, the parallel and antiparallel--spin contribution
to the plasma mode is the same. Moreover, to build up
model functions for the spin--resolved $S_{\sigma_1\sigma_2}$, it's
crucial to include the following property of the so--called
magnetic structure factor $S_{\uparrow\uparrow}
-S_{\uparrow\downarrow}$:
\begin{equation}
S_{\uparrow\uparrow}-S_{\uparrow\downarrow}\big|_{k\to 0} =
\frac{3}{4}k-\frac{k^3}{16}+{\cal O}(k^4),
\label{Sm}
\end{equation}
where the scaled variable $k=q/q_F$ has been used.
Eq.~(\ref{Sm}) is valid in the framework of the random--phase
approximation\cite{histo}
(RPA) and can be obtained from a series expansion
of $S_{\uparrow\downarrow}^{\rm RPA}(k)$
near $k=0$ (Ref.~\CITE{Ueda}), and from the corresponding expansion of the total
$S^{\rm RPA}$ (see for instance Ref.~\CITE{PW_nxcHD}).
Since in the $k\to 0$ limit
RPA is exact,\cite{histo,PW_nxcHD,Gell-Mann,RPA}
we expect Eq.~(\ref{Sm}) to
hold for the exact structure factor as well.
From Eqs.~(\ref{plasmon}) and~(\ref{Sm}) we can write the
small--$k$ expansion of $S_{\sigma_1\sigma_2}$:
\begin{eqnarray}
S_{\uparrow\downarrow}\big|_{k\to 0} & = & -\frac{3}{8}k+
\frac{q_F^2\, k^2}{4\, \omega_p(r_s)}+\frac{k^3}{32}+{\cal O}(k^4) \label{Sudq0}\\
S_{\uparrow\uparrow}\big|_{k\to 0} & = & \frac{3}{8}k+
\frac{q_F^2\, k^2}{4\, \omega_p(r_s)}-\frac{k^3}{32}+{\cal O}(k^4). \label{Suuq0}
\end{eqnarray}
\subsection{Correlation energy}
\label{sub_EHD}
The electron--electron potential energy is, as known, given
by the sum of repulsive two--body Coulomb terms:
\begin{equation}
U=\frac{1}{2} \sum_{i \ne j}^N \frac{1}{|{\bf r}_i-{\bf r}_j|}.
\label{pot_en}
\end{equation}
Its ground--state expectation value (per electron), in a homogenous
electron gas
of density $r_s$, is given by the following integral over the
pair--correlation function:
\begin{equation}
\langle U\rangle_{r_s} = \frac{3}{2\,\alpha^2 \,r_s} \int_0^{\infty}
\left[ g(\rho;r_s)-1 \right]\,\rho\,d\rho.
\label{<U>}
\end{equation}
By the virial theorem~\cite{virial} and the
usual definition of the correlation energy $\epsilon_c$
as total electronic energy minus Hartree--Fock energy, we have:
\begin{equation}
\langle U \rangle_{r_s}=-\frac{3\,q_F}{4\,\pi}+\frac{1}{r_s}\frac{d}{d r_s}
\left[r_s^2\,\epsilon_c(r_s)\right]
\label{virial}
\end{equation}
Putting together Eqs.~(\ref{<U>}) and~(\ref{virial}) one obtains the
known relation between $g(\rho;\,r_s)$ and the exchange and
correlation energy:
\begin{equation}
\epsilon_{xc}=-\frac{3q_F}{4\pi}+\epsilon_c=\frac{3}{2 \alpha^2 r_s^2}
\int_0^{\infty}d\rho\,\rho\int_0^{r_s} dr_s'\left[g(\rho;\,r_s')-1\right]
\label{eq_excfromg}
\end{equation}
The same relation can
be obtained in a more general way~\cite{gunlun} by the
Hellmann--Feynman theorem and the
coupling--constant average of $g(\rho;\,r_s)$,
which, for the homogeneous system is just the average over $r_s$:
\begin{equation}
\overline{g}(\rho;\,r_s)= \frac{1}{r_s}\int_0^{r_s}g(\rho;\,r_s')\,dr_s'.
\label{g_average}
\end{equation}
The function $\overline{g}(\rho;\,r_s)$ is directly related to the exchange and
correlation hole\cite{gunlun,PerdewWang} of the electron gas.

We have recalled these relations because we will later check our analytical
expressions for $g(\rho;\,r_s)$ against available energy data, and also
because, among other constraints, we want our functional
form of $g(\rho;r_s)$ to be consistent with the high--density limit of 
$\epsilon_c(r_s)$:
\begin{equation}
\epsilon_c(r_s \to 0) = A\,\ln r_s+B+C\,r_s\,\ln r_s+D\,r_s
\label{Ec_HD}
\end{equation}
where $A$, $B$, $C$ and $D$ are known
constants,\cite{Gell-Mann,PWEc,HDEc,WangPerdewEcHD,Hoffman,Onsager,DHD}
$A=(1-\ln 2)/\pi^2$, $B=-0.0469205$, $C=0.0092292$, $D=-0.01$,
and the next leading term is ${\cal O}(r_s^2\ln r_s)$.

\section{General strategy}
\label{sec_strat}
We study the antiparallel and parallel--spin
correlation functions both in real and reciprocal space and we split them,
as usually, into exchange and correlation according to:
\begin{eqnarray}
g_{\uparrow\downarrow}(\rho;r_s)  & = & 1+g^c_{\uparrow\downarrow}(\rho;r_s) \\
g_{\uparrow\uparrow}(\rho;r_s) & = & g_{ex}(\rho)+
g^c_{\uparrow\uparrow}(\rho;r_s) \\
S_{\uparrow\downarrow}(k;r_s) & = & S^c_{\uparrow\downarrow}(k;r_s)
\label{eq_sepSud}\\
S_{\uparrow\uparrow}(k;r_s) & = & S_{ex}(k)+S^c_{\uparrow\uparrow}(k;r_s)
\label{eq_sepSuu}
\end{eqnarray}
where the exchange functions, given by the Hartree--Fock
approximation, are equal to:
\begin{eqnarray}
g_{ex}(\rho) & = & 1-9\left(\frac{\sin\rho-\rho\cos\rho}{\rho^3}
\right)^2 \label{gHF} \\
S_{ex}(k) & = & \cases{ 3 k/4-k^3/16 &for $k\le 2$ \cr
                1  &for $k>2$\cr}
\label{SHF}
\end{eqnarray}
and our model only concerns the correlation part.
Putting together Eqs.~(\ref{eq_Stot}), (\ref{plasmon}),
(\ref{eq_sepSud}), (\ref{eq_sepSuu}) and~(\ref{SHF}) one finds
a well--known result: in the total $S=S_{ex}+S^c_{\uparrow\downarrow}+
S^c_{\uparrow\uparrow}$,
the linear term of $S_{ex}(k)$, $3 k/4$,
which dominates its small--$k$ behavior (and corresponds to
a large--$\rho$ leading
term $\propto 1/\rho^4$ of $g_{ex}$) exactly cancels
the small--$k$ leading term of the correlation part
$S^c_{\uparrow\downarrow}+S^c_{\uparrow\uparrow}$.
This property has been incorporated in several previous
functional forms for the total $g$, as the widely used
Perdew--Wang\cite{PerdewWang} model (hereafter PW) where,
however, the $k^2$
coefficient in the small--$k$ expansion
of $S(k)$ is slightly different
from the exact one [Eq.~(\ref{plasmon})],
because of spurious $k^2$ contributions
from their $\langle g_x\rangle$ [Eq.~(19) of
Ref.\CITE{PerdewWang}] and from their short--range
part of $g_c$ [Eq.~(37) of Ref.\CITE{PerdewWang}].

The $k\to 0$ limit of Eq.~(\ref{Sm})
seems, instead, to be less known: even the best--to--date
spin--resolved PW\cite{PerdewWang} model
does not incorporate such a non--trivial
analytic property, which can alternatively be expressed as
$S^c_{\uparrow\uparrow}$ beeing identical to $S^c_{\uparrow\downarrow}$
in the  the small--$k$ limit and
corresponds to a visible feature of the magnetic structure factor
(see Sec.~\ref{sec_results}).
Our goal is to produce simple and practical analytical functional forms for
$S^c_{\uparrow\downarrow}(k;r_s)$ and $S^c_{\uparrow\uparrow}(k;r_s)$
[and hence
$g^c_{\uparrow\downarrow}(\rho,r_s)$ and $g^c_{\uparrow\uparrow}(\rho,r_s)$]
which
satisfy all the physical properties of Sec.~\ref{sec_properties} and have
enough
variational flexibility to accurately interpolate the QMC data of Ortiz, Harris
and Ballone.\cite{OBH}

To do this, let's start with a few simple considerations about
spherical Fourier transforms: $S^c_{\sigma_1\sigma_2}(k)$ and
$g^c_{\sigma_1\sigma_2}(\rho)$ are related to one another
by an integration like
Eq.~(\ref{Sud_def}). The function
$\sin(k\rho)/k\rho$ is an even function, i. e. its odd derivatives
in $k=0$ (or $\rho=0$) are all equal to zero. However, the small--$\rho$
and the small--$k$ properties of $g$ and $S$ tell us that they must
have non--zero odd derivatives in $\rho=0$ and $k=0$. This
is achieved if (and only if), as the integration variable goes to infinity, the 
integrand goes to zero slowly enough as to avoid
absolute convergence, so that
differentiation within the integral sign
is not allowed. It is easy to establish a connection
between the large--$k$ (large--$\rho$) behavior of $S$ ($g$) and the
odd derivatives in $\rho=0$ ($k=0$) of $g$ ($S$): a derivative
of $g$ in $\rho=0$ of order $2n+1$ corresponds, in $S$, to a large--$k$ term
$\propto 1/k^{2n+4}$ and viceversa. This simple relation was
used in Ref.\CITE{Kimball2} to obtain the large--$k$
expansion of $S_{\sigma_1\sigma_2}$ from the cusp conditions
of Eqs.~(\ref{cusp_ud}) and~(\ref{cusp_uu}).
These elementary considerations lead us to write down a very
simple functional form $S^c$ in reciprocal space which
automatically has the exact small--$k$ and large--$k$
behavior. Its spherical
Fourier transform $g^c$ is analytic and closed--form, consisting
of the same kind of functions used in
reciprocal space. We thus have an equally simple expression for
$S^c$ and $g^c$.

We begin by studying the antiparallel--spin part, and do it in
several steps. First (\ref{sec_funcud}) we choose our
functional form. Then (\ref{sec_construd}) we impose to it
the properties of Sec.~\ref{sec_properties}. At this point
we are left with 6 free parameters, which, independently for
each available $r_s$, are used to accurately fit the QMC data
both in real and reciprocal space, as done in Ref.~\CITE{OB}. 
In our case, however, the $r_s$ dependence of each
of the 6 optimal parameters turns out to be both regular and monotonic.
We then try to represent each of them as a simple function
of $r_s$ in such a way that (i)  as $r_s \to 0$ the exact 
high--density expansion of 
the correlation energy [Eq.~(\ref{Ec_HD})] is recovered (\ref{sub_HDud}),
and (ii) for finite $r_s\le 10$ an optimal global fit of all
the QMC data\cite{OBH} is obtained (\ref{sec_fitQMC}).
We apply the same strategy to the parallel--spin part (\ref{sec_uu}). 
Besides the excellent quality of the final fits of $g$ and $S$, we
see that even the resulting correlation energy, not 
targeted by our fitting
strategy except at $r_s \to 0$, turns out to be in
good agreement (within $5\%$) with the corresponding QMC results\cite{OBH}
at any $r_s$.
We compare our correlation energy with the most popular interpolation
formulae and we discuss their relative
efficiency in fitting the new QMC
energies.\cite{OBH}
\section{Antiparallel spins}
\label{sec_ud}
\subsection{Functional form}
\label{sec_funcud}
In reciprocal space our functional form is simply written as:
\begin{eqnarray}
S^c_{\uparrow\downarrow}(k;r_s) & = & \exp\left[-b^{\uparrow\downarrow}(r_s)
\,k\right]\,
\sum_{n=1}^6 c_n^{\uparrow\downarrow}(r_s)\,k^n+ \nonumber \\
& & \frac{\alpha_6^{\uparrow\downarrow}(r_s)\,k^8+
\alpha_4^{\uparrow\downarrow}(r_s)\,k^{10}}{\left[(a^{\uparrow\downarrow})^2
+k^2\right]^7};
\label{Sud}
\end{eqnarray}
as mentioned, the corresponding $g^c_{\uparrow\downarrow}$
amounts to a linear combination of the same kind of
functions\cite{serieallucinante} in real space
[see App.~\ref{app_g}, Eq.~(\ref{gud})].
Two types of functions appear in
Eq.~(\ref{Sud}): the first one, an exponential cut--off times
a truncated power series,
fully characterizes the long--wavelength behavior of $S$, while
the second one entirely determines its large--$k$ expansion. The
leading term as $k\to \infty$ is of order $k^{-4}$, as exactly known
from the cusp condition;\cite{Kimball2,Kimball1}
in real space the short--range
behavior is thus entirely determined by the parameter $\alpha_4^{\uparrow
\downarrow}(r_s)$:
\begin{eqnarray}
\frac{\partial}{\partial \rho}g_{\uparrow \downarrow}(\rho;r_s)
\biggr|_{\rho \to 0} & = & -\frac{3\pi}{4}\alpha_4^{\uparrow
\downarrow}(r_s) \label{gudprime} \\
g_{\uparrow \downarrow}(\rho=0;r_s) & = & -\frac{3\pi}{4} q_F(r_s)
\alpha_4^{\uparrow\downarrow}(r_s). \label{g0_alpha4}
\end{eqnarray}
\subsection{Physical constraints}
\label{sec_construd}
The $k\to 0$ conditions of Subsec.~\ref{sec_q0prop} are easily imposed:
\begin{eqnarray}
c_1^{\uparrow\downarrow} & = & -\frac{3}{8} \label{cud1} \\
c_2^{\uparrow\downarrow} & = & b^{\uparrow\downarrow}
c_1^{\uparrow\downarrow}+\frac{q_F^2}{4\omega_p}
\label{cud2} \\
c_3^{\uparrow\downarrow} & = & (b^{\uparrow\downarrow})^2\frac{c_1^
{\uparrow\downarrow}}{2}+b^{\uparrow\downarrow}\frac{q_F^2}{4\omega_p}+
\frac{1}{32}. \label{cud3}
\end{eqnarray}
The cusp condition of Eq.~(\ref{cusp_ud}) fixes a simple relation
between $\alpha_6^{\uparrow\downarrow}$ and the other parameters:
\begin{eqnarray}
\alpha_6^{\uparrow\downarrow} & = & (a^{\uparrow\downarrow})^3
\Biggl\{\alpha_4^{\uparrow\downarrow}\left(\frac{11}{a^{\uparrow\downarrow}}-
\frac{512}{21}q_F\right)- \nonumber \\
& & \frac{2048}{21\pi}\left[\frac{1}{3}+\sum_{n=1}^6c_n^{\uparrow\downarrow}
\frac{(n+2)!}{(b^{\uparrow\downarrow})^{n+3}}\right]\Biggr\}
\end{eqnarray}
After imposing all the $\uparrow\downarrow$ physical conditions,
our model [Eq.~(\ref{Sud})] is
left with 6 free parameters: the two
exponential cut--offs ($a^{\uparrow\downarrow}$
in real space and $b^{\uparrow\downarrow}$ in reciprocal space),
the parameter $\alpha_4^{\uparrow\downarrow}$,
which determines the short--range behavior of $g(r)$, and the 3 linear
parameters $c_4^{\uparrow\downarrow}$, $c_5^{\uparrow\downarrow}$ and
$c_6^{\uparrow\downarrow}$, which will be used to further 
increase the variational flexibility and fit the
numerical $g_{\uparrow\downarrow}$
obtained from QMC simulations.\cite{OBH}
The dependence of these free parameters on $r_s$ will be determined
according to the strategy summarized in Sec.~\ref{sec_strat} and detailed
in the following Subsec.~\ref{sub_HDud} and Sec.~\ref{sec_uu}.
\subsection{High--density expansion}
\label{sub_HDud}
As anticipated in Subsec.~\ref{sub_EHD}, we want our
pair--correlation function such that its $\uparrow\downarrow$ and
$\uparrow\uparrow$ contributions automatically fulfill the
high--density limit of the correlation energy. We thus
fix the $r_s \to 0$ limit of our free parameters
by means of
Eqs.~(\ref{<U>}), (\ref{virial}) and (\ref{Ec_HD}).
Our antiparallel contribution to the expectation value of
the potential energy $U=U_{\uparrow\downarrow}+U_{\uparrow\uparrow}$
is simply given by:
\begin{equation}
U_{\uparrow\downarrow}=\frac{q_F}{\pi}\sum_{n=1}^6
\frac{c_n^{\uparrow\downarrow}n!}{(b^{\uparrow\downarrow})^{n+1}}+
\frac{q_F}{2048}\left[\frac{7\alpha_6^{\uparrow\downarrow}}{(a^
{\uparrow\downarrow})^5}+\frac{21\alpha_4^{\uparrow\downarrow}}{(a^
{\uparrow\downarrow})^3}\right].
\label{Ucud}
\end{equation}
In the high--density limit, the correlation--energy constraint 
of Eq.~(\ref{Ec_HD}) translates into the following condition on
$\langle U_{\uparrow \downarrow} \rangle_{r_s}$:
\begin{equation}
\langle U_{\uparrow \downarrow}\rangle_{r_s\to 0} =
2\,A_{\uparrow \downarrow}\,\ln r_s+
\left(A_{\uparrow \downarrow}+2\,
B_{\uparrow \downarrow}\right)+{\cal O}(r_s\,\ln r_s)
\label{potud_HD}
\end{equation}
where, comparing to Eq.~(\ref{Ec_HD}),
 $A_{\uparrow \downarrow}$ is simply given\cite{WangPerdewEcHD} by
$\frac{1}{2}A$. To determine
$B_{\uparrow \downarrow}$ we recall that the constant $B$
in Eq.~(\ref{Ec_HD}) is the sum of two contributions: a second--order
exchange term, $B^{(2)}_{exc}$, which only concerns the
$\uparrow\uparrow$ part,
and a direct term, $B_d$, which is, instead, equally
split (in the unpolarized gas) between $\uparrow\uparrow$ and
$\uparrow\downarrow$. Hence,
$B_{\uparrow \downarrow}=B_d/2$. Both $B^{(2)}_{exc}$ and $B_d$
have been evaluated exactly.\cite{Hoffman,Onsager}

Provided that the 3 linear parameters $c_4^{\uparrow \downarrow}$,
$c_5^{\uparrow \downarrow}$ and $c_6^{\uparrow \downarrow}$ remain
finite as $r_s \to 0$, the exact high--density limit of
Eq.~(\ref{potud_HD}) amounts to the following conditions:
\begin{eqnarray}
\alpha_4^{\uparrow \downarrow}(r_s\to 0) & = & \frac{
1+k_1 r_s \ln r_s+k_2 r_s+ {\cal O}(r_s^2 \ln r_s)}
{-3\pi q_F/4} \label{alf4HD} \\
b^{\uparrow \downarrow}(r_s \to 0) & = & \left(\frac{4}{9\pi}
\right)^{1/3}\pi\sqrt{\frac{3}{r_s}}+{\cal O}(r_s^0) \label{budHD} \\
a^{\uparrow \downarrow}(r_s \to 0) & = & {\rm const.} +{\cal O}(r_s)
\equiv a^{\uparrow\downarrow}+{\cal O}(r_s) \label{audHD}
\end{eqnarray}
where $k_1$ and $k_2$ depend on $A_{\uparrow \downarrow}$,
$B_{\uparrow \downarrow}$ and $a^{\uparrow\downarrow}$:
\begin{eqnarray}
k_1 & = & \frac{18\pi (a^{\uparrow\downarrow})^2}{\alpha}
A_{\uparrow\downarrow}\label{k1ud} \\
k_2 & = & \frac{729}{64}\frac{(a^{\uparrow\downarrow})^2}{
\alpha^4}-\frac{21}{64}\frac{1}{a^{\uparrow\downarrow}\alpha}+
\nonumber \\
& & \frac{9 (a^{\uparrow\downarrow})^2\pi}{2\alpha}
(A_{\uparrow\downarrow}+2 B_{\uparrow\downarrow}).
\label{k2ud}
\end{eqnarray}
Inserting Eq.~(\ref{alf4HD}) into Eq.~(\ref{g0_alpha4}) we see that
in our model, once
the high--density expansion of $\epsilon_c(r_s)$
is fixed, the $r_s \to 0$ limit of $g(\rho=0;r_s)$ is also fixed
to the form $1+C r_s\ln r_s+{\cal O}(r_s)$.
The corresponding exact form\cite{KimballHD} is,
up to orders $r_s^2$,
$1+C_1r_s+C_2r_s^2\ln r_s$, thus slightly different from
ours. Evidently,
the simple functional form of Eq.~(\ref{Sud}) does not
correctly describe the short--range behavior of the Coulomb hole at
very high densities. It is worthwhile
to point out that this contradiction is due to the
exponential cut--off times a truncated power series in real
space [Eq.~(\ref{gud})], and emerges when the cusp condition
of Eq.~(\ref{cusp_ud}) is imposed to it. 
The relevance of this limitation, which only concerns
densities $r_s\lesssim 0.1$, will be discussed in next
Sec.~\ref{sec_results}.
\section{Parallel spins}
\label{sec_uu}
\subsection{Functional form}
For the correlation part of the $\uparrow\uparrow$ pair--distribution
function we apply the same strategy used for the antiparallel--spin
case. In reciprocal space we thus have:
\begin{eqnarray}
S^c_{\uparrow\uparrow}(k;r_s) & = & \exp\left[-b^{\uparrow\uparrow}(r_s)\,k
\right]\,
\sum_{n=1}^6 c_n^{\uparrow\uparrow}(r_s)k^n+ \nonumber \\
& & \frac{\alpha_{10}^{\uparrow\uparrow}(r_s)k^8+
\alpha_8^{\uparrow\uparrow}(r_s)k^{10}+\alpha_6^{\uparrow\uparrow}(r_s)
k^{12}}{\left[(a^{\uparrow\uparrow})^2+k^2\right]^9}
\label{Suu}
\end{eqnarray}
which again corresponds, in real space, to a linear combination of the same kind
of functions\cite{serieallucinante} [see App.~\ref{app_g}, Eq.~(\ref{guu})].
The long--wavelength term has the same form as
the $\uparrow\downarrow$ part. The
large--$k$ term describes the short--range behavior
of $g_{\uparrow\uparrow}$: the cusp condition of Eq.~(\ref{cusp_uu})
tells us that, as $k \to \infty$,
the leading term of $S^c_{\uparrow\uparrow}$ 
must be of order $k^{-6}$. With respect to the
$\uparrow\downarrow$ case, one more parameter is needed for 
the large--$k$ term to
satisfy the Pauli principle.
As in the antiparallel--spin case, the short--range properties
of $g_{\uparrow\uparrow}$ are characterized by the
$\alpha_6^{\uparrow\uparrow}$ parameter:
\begin{eqnarray}
\frac{\partial^3}{\partial \rho^3}g_{\uparrow \uparrow}(\rho;r_s)
\biggr|_{\rho \to 0} & = & \frac{3\pi}{8}\alpha_6^{\uparrow
\uparrow}(r_s) \label{guuthird} \\
\frac{\partial^2}{\partial \rho^2}g_{\uparrow \uparrow}(\rho;r_s)
\biggr|_{\rho \to 0} & = & \frac{\pi}{4} q_F(r_s)
\alpha_6^{\uparrow\uparrow}(r_s). \label{guu20_alpha6}
\end{eqnarray}
\subsection{Physical constraints}
The small--$k$ properties imply, for the $\uparrow\uparrow$ case, identical
constraints as for the
$\uparrow\downarrow$ case [see Eqs.~(\ref{cud1}), (\ref{cud2})
and~(\ref{cud3})].

The Pauli principle and the cusp condition of Eqs.~(\ref{Pauli})
and~(\ref{cusp_uu}) fix the dependence of $\alpha_8^{\uparrow
\uparrow}$ and $\alpha_{10}^{\uparrow\uparrow}$ on the remaining parameters:
\begin{eqnarray}
\alpha_8^{\uparrow\uparrow} & = & \frac{2048}{3\pi}(a^{\uparrow\uparrow})^5
\sum_{n=1}^6\frac{c_n^{\uparrow\uparrow}}{(b^{\uparrow\uparrow})^{n+3}}
\left[(n+2)!-\frac{5(n+4)!}{(a^{\uparrow\uparrow}b^{\uparrow\uparrow})^2}
\right] \nonumber \\
& & +\frac{4096}{33\pi}(a^{\uparrow\uparrow})^3-\alpha_6^{\uparrow\uparrow}
(a^{\uparrow\uparrow})^3\left(\frac{2560q_F}{33}+
\frac{26}{a^{\uparrow\uparrow}}
\right) \label{alf8uu} \\
\alpha_{10}^{\uparrow\uparrow} & = & \frac{2048}{3\pi}
(a^{\uparrow\uparrow})^7\sum_{n=1}^6\frac{c_n^{\uparrow\uparrow}}
{(b^{\uparrow\uparrow})^{n+3}}\left[
\frac{(n+4)!}{(a^{\uparrow\uparrow}b^{\uparrow\uparrow})^2}-
\frac{13}{3}(n+2)!\right]
\nonumber \\
& & -\frac{4096}{15 \pi}(a^{\uparrow\uparrow})^5+
\frac{\alpha_6^{\uparrow\uparrow}}{3}(a^{\uparrow\uparrow})^5\left(
\frac{143}{a^{\uparrow\uparrow}}+512 q_F\right).
\label{alfa10uu}
\end{eqnarray}
The cusp condition of Eq.~(\ref{cusp_uu}) is not included in the PW model.
As in the antiparallel--spin case, we have 6 free parameters: the exponential
cut--off in real space, $a^{\uparrow\uparrow}$, the exponential cut--off
in reciprocal space, $b^{\uparrow\uparrow}$, 
the $\alpha_6^{\uparrow\uparrow}$
parameter, which determines the short--range behavior
of $g_{\uparrow\uparrow}$, and the three linear parameters
$c_4^{\uparrow\uparrow}$, $c_5^{\uparrow\uparrow}$ 
and $c_6^{\uparrow\uparrow}$, which are
used to fit the oscillatory behavior of
$g_{\uparrow\uparrow}$.
\subsection{High--density expansion}
The contribution to the expectation value of the potential energy
due to the correlation part of our $g_{\uparrow\uparrow}$
($U_{\uparrow\uparrow}=-3q_F/4\pi+U^c_{\uparrow\uparrow}$) is:
\begin{eqnarray}
U_{\uparrow\uparrow}^c & = & \frac{q_F}{\pi}\sum_{n=1}^6
\frac{c_n^{\uparrow\uparrow} n!}{(b^{\uparrow\uparrow})^{n+1}}+
\frac{q_F}{65536}\biggl[\frac{35\,\alpha_{10}^{\uparrow\uparrow}}{(
a^{\uparrow\uparrow})^9}+\frac{45\,\alpha_8^{\uparrow\uparrow}}{
(a^{\uparrow\uparrow})^7}+ \nonumber \\
& & +
\frac{99\,\alpha_6^{\uparrow\uparrow}}{(a^{\uparrow\uparrow})^5}
\biggr] \label{pot_uu}
\end{eqnarray}
As $r_s\to 0$, the condition on $U_{\uparrow\uparrow}^c$ is identical
to Eq.~(\ref{potud_HD}),
where\cite{WangPerdewEcHD} $A_{\uparrow\uparrow}=\frac{1}{2}A$, and
$B_{\uparrow\uparrow}=B_{exc}^{(2)}+\frac{1}{2}B_d$. As in the
$\uparrow\downarrow$ case, the exact $r_s\to 0$ expansion
of $U_{\uparrow\uparrow}^c$ implies for the two exponential
cut--offs in real ($a^{\uparrow\uparrow}$) and reciprocal
space ($b^{\uparrow\uparrow}$) identical conditions as
Eqs.~(\ref{budHD}) and~(\ref{audHD}). For
$\alpha_6^{\uparrow\uparrow}$ the condition is similar
to Eq.~(\ref{alf4HD}):
\begin{eqnarray}
\alpha_6^{\uparrow\uparrow}(r_s\to 0) & = & \frac{4}{\pi q_F}
\biggl[\frac{2}{5}+p_1 r_s \ln r_s+p_2 r_s + \nonumber \\
& & +{\cal O}(r_s^2\ln r_s)\biggr], \label{alf6uuHD}
\end{eqnarray}
where $p_1$ and $p_2$ depend on $A_{\uparrow\uparrow}$,
$B_{\uparrow\uparrow}$ and $a^{\uparrow\uparrow}$ through
equations similar to Eqs.~(\ref{k1ud}) and~(\ref{k2ud}).
From Eqs.~(\ref{guu20_alpha6}) and~(\ref{alf6uuHD}) one can see
that, as expected, when $r_s\to 0$, $g_{\uparrow\uparrow}''(\rho=0)$
goes to the Hartree--Fock value $2/5$. As in the antiparallel--spin
case, the high--density limit of the correlation energy
fixes the $r_s\to 0$ expansion of $g_{\uparrow\uparrow}''(\rho=0;r_s)$.
The exact $g_{\uparrow\uparrow}''(\rho=0;r_s\to 0)$ should have the
form\cite{Kimball2} $2/5+{\cal O}(r_s)$, while our functional
form gives $2/5+{\cal O}(r_s \ln r_s)$. Again, we find that in
real space the simple exponential cut--off times a truncated power
series [Eq.~(\ref{guu})] does not correctly describe
the short range Coulomb interactions
at very high densities.
\section{Fit to QMC data}
\label{sec_fitQMC}
For each available density
in the range $0.8 \le r_s \le 10$  (i.e. $r_s=0.8$,
1, 2, 3, 4, 5, 8 and 10) we performed a best fit
of the 6 free parameters to the
QMC data,\cite{OBH}
separately for the $\uparrow\downarrow$ and
the $\uparrow\uparrow$ parts. The $r_s$
dependence of the parameters turns out to be quite smooth and
monotonic and well described
by the following functional forms ( which also take into account
the exact high--density limit of Eqs.~(\ref{alf4HD})--(\ref{audHD})
and~(\ref{alf6uuHD})
and guarantee the exact low--density expansion of the resulting
correlation energy\cite{cepald,PZ,PWEc,Vosko,Aguilera}
$d_1r_s^{-1}+d_2r_s^{-3/2}$):
\begin{eqnarray}
\alpha_4^{\uparrow\downarrow}(r_s) & = &-
\frac{4\left[1-k_1(a^{\uparrow\downarrow})r_s \ln\left(
1+\tilde{k_2}(a^{\uparrow\downarrow})/r_s\right)\right]}{3\pi q_F\left(1+
k_3r_s^2\right)} \label{alfa4ud} \\
\alpha_6^{\uparrow\uparrow}(r_s) & = & \frac{8\left[1-p_1
(a^{\uparrow\uparrow})r_s \ln\left(
1+p_2(a^{\uparrow\uparrow})/r_s\right)\right]}{5\pi q_F\left(1+
p_3r_s^2\right)} \label{alfa6uu} \\
a^{\sigma_1\sigma_2}(r_s) & = & a^{\sigma_1\sigma_2} \label{audauu}\\
b^{\sigma_1\sigma_2}(r_s) & = & \left(\frac{4}{9\pi}
\right)^{1/3}\pi\sqrt{\frac{3}{r_s}}+b_1^{\sigma_1\sigma_2}
\label{budbuu} \\
c_n^{\sigma_1\sigma_2}(r_s) & = &
\frac{\lambda_n^{\sigma_1\sigma_2}+\gamma_n^{\sigma_1\sigma_2}
r_s}{1+r_s^{3/2}}\;\;\;\;\;\;\;\; n=4,5,6
\label{cudcuu456}
\end{eqnarray}
where $k_1(a^{\uparrow\downarrow})$ is given by Eq.~(\ref{k1ud}),
and $\tilde{k_2}(a^{\uparrow\downarrow})$, $p_1(a^{\uparrow\uparrow})$
and $p_2(a^{\uparrow\uparrow})$ are equal to:
\begin{eqnarray}
\tilde{k_2}(a^{\uparrow\downarrow}) & = & \exp\biggl[\frac{7}
{384\pi(a^{\uparrow\downarrow})^3A_{\uparrow\downarrow}}-\frac{81}
{128\pi\alpha^3A_{\uparrow\downarrow}}
\nonumber \\ & & -\frac{B_{\uparrow\downarrow}}
{A_{\uparrow\downarrow}}-\frac{1}{2}\biggr] \label{costud}\\
p_1(a^{\uparrow\uparrow}) & = & \frac{33\pi A_{\uparrow\uparrow}
(a^{\uparrow\uparrow})^4}{\alpha} \label{p1uu}\\
p_2(a^{\uparrow\uparrow}) & = & \exp\biggl[\frac{7}{960\pi
(a^{\uparrow\uparrow})^5A_{\uparrow\uparrow}}
-\frac{81}{128\pi\alpha^3
A_{\uparrow\uparrow}} \nonumber \\ & & -\frac{B_{\uparrow
\uparrow}}{A_{\uparrow\uparrow}}-\frac{1}{2}\biggr]. \label{p2uu}
\end{eqnarray}
The 9 constants for $\uparrow\downarrow$ and
the 9 constants for $\uparrow\uparrow$
have been fixed by a two--dimensional best fit to
the QMC data in real and reciprocal space (9368+9368 data points).
The efficiency of our interpolation scheme has been tested by
performing preliminary fits in wich some of the available $r_s$ were
not included and then verifying that the interpolated $g$ and $S$ were in good
agreement with the corresponding QMC quantities for
the excluded $r_s$. Since this was always the case, we
included all the available $r_s$ in order to have optimal values for our
final parameters. We thus expect our $g$ and $S$ to be very reliable and
accurate in the whole density range $r_s \in [0.8,10]$.
The optimal 9 parameters which define our best antiparallel--spin 
model are:
$a^{\uparrow\downarrow}=0.838$,
$k_3=0.141$, $b_1^{\uparrow\downarrow}=3.27$,
$\lambda_4^{\uparrow\downarrow}=-78$, $\gamma_4^{\uparrow\downarrow}
=28$, $\lambda_5^{\uparrow\downarrow}=216$,
$\gamma_5^{\uparrow\downarrow}=-124$,
$\lambda_6^{\uparrow\downarrow}=-140$, $\gamma_6^{\uparrow\downarrow}
=55$, and the other 9 parameters for the parallel--spin part are:
$a^{\uparrow\uparrow}=1.32$,
$p_3=0.015$, $b_1^{\uparrow\uparrow}=3.47$,
$\lambda_4^{\uparrow\uparrow}=98$, $\gamma_4^{\uparrow\uparrow}
=-36$, $\lambda_5^{\uparrow\uparrow}=-295$,
$\gamma_5^{\uparrow\uparrow}=74$,
$\lambda_6^{\uparrow\uparrow}=170$, $\gamma_6^{\uparrow\uparrow}
=-13$.
\section{Results in real and reciprocal space}
\label{sec_results}
After fixing the $9+9$ parameters which fully specify our model,
we are now ready to
present, in Fig.~\ref{fig_g}, our real--space pair--correlation
function $g_{\sigma_1\sigma_2}$, shown as a solid line, as a function
of the scaled variable $r/r_s$.  This is done for the 8 values of
$r_s$ for which QMC results\cite{OBH}, shown as solid dots, were
available.  The best-to-date model correlation function
of Perdew--Wang\cite{PerdewWang} (PW) is also shown for comparison as a
dashed line. 
Perdew and Wang\cite{PerdewWang} interpolated the total pair-correlation
function $g(r)$ between its short--range limit, dominated by the on-top
value and cusp, and the nonoscillatory part of its long--range limit.
Their interpolation, controlled by normalization and energy integrals,
agreed with older spin-unresolved QMC data\cite{cepald}.  
They only needed the total $g(r)$
for construction of the generalized gradient approximation,\cite{GGA}
however, they also made an estimate for the spin resolution of $g$,
using scaling relations that preserve the normalization integrals but
are exact only for the exchange contribution.

Our new expression, explicitly constructed to fit spin--resolved numerical
correlation functions, follows the QMC data\cite{OBH} better (low $r_s$) 
or much better (medium and high $r_s$)
than the corresponding PW model, whose performance with
respect to the new QMC data\cite{OBH} becomes reasonable only
after summing the two contributions and going back
to the total, spin--unresolved version (not shown).  
This can be guessed from the
fact that for $r_s \ge 2$, where the discrepancies become clearly
visible, they generally have opposite signs: both the up--up and
the up--down correlations are larger (i. e. less close to one) than they
should. This is
due to the fact that the PW estimate for the $\uparrow\uparrow$
part is a simple rescaling of the pair--correlation function of
the fully polarized gas,\cite{newPW} while in the 
unpolarized case correlations
are dominated by $\uparrow\downarrow$ interactions (see for instance
Ref.~\CITE{histo}). 
Like the PW model, our pair--correlation function
breaks down for $r_s>10$: for very low densities $g$ tends to
become negative at small $\rho$. This is probably due to
the limited variational flexibility of the model, which in this
low--density regime cannot at the same time fulfill the cusp
conditions at $\rho=0$ and reproduce a flatter and flatter, 
yet non--negative $g$ for $\rho\gtrsim 0$. As we shall see in the next 
Sec.~\ref{sec_correnergy}, such a breakdown has no impact on the  
resulting correlation energy, which is an integral of $g$ and
remains accurate at any $r_s$.

We compare in Fig.~\ref{fig_g0} our $g_{\uparrow\downarrow}(\rho=0;r_s)$
(solid line)
to the Yasuhara\cite{Yas} electron--electron ladder approximation (dashed
line). Built up to interpolate the QMC
data, our
$g_{\uparrow\downarrow}(\rho=0;r_s)$ is larger than
the Yasuhara result for $r_s \gtrsim 0.5$,
as expected from Fig.~\ref{fig_g}, where
the discrepancy between the short--range behavior of the QMC
data and the PW model (which by construction follows
the Yasuhara approximation) is clearly visible.
In the inset the corresponding high--density expansions are
shown, together with the exact limit\cite{KimballHD} (dots), which,
as anticipated in Subsec.~\ref{sub_HDud}, is not fulfilled by our 
$g_{\uparrow\downarrow}(\rho=0;r_s)$. Rather than giving up our
exact limit of $\epsilon_c(r_s)$ for $r_s\to 0$ or trying
to fulfill both the $\epsilon_c$ and the $g_{\uparrow\downarrow}(0)$
limits, we have preferred to accept a slight discrepancy
of $g(0)$ and keep our functional form as described up to now: in
our experience the collateral complications, at least within 
our functional form, were not worth the effort. 
Because of this limitation, our pair--correlation function
does not fulfill the high--density limit of $g^c/r_s$ recently
computed by Rassolov~{\it et~al}.\cite{Ras}

In Fig.~\ref{fig_S} we report the total static structure factor
$S_{\uparrow\uparrow}+S_{\uparrow\downarrow}$ and the magnetic
structure factor $S_{\uparrow\uparrow}-S_{\uparrow\downarrow}$ for
the same 8 values of $r_s$ as in Fig.~\ref{fig_g}. Again, our
model is shown as solid lines, the QMC data\cite{OBH} as dots
and the PW model\cite{PerdewWang} as dashed lines. Our combination
of analytic constraints and fitting procedure nicely
interpolates the QMC data, filtering out their noise. In reciprocal
space it becomes clear that the long--range of the PW spin--resolved model
is not exact.
Moreover, as said in Sec.~\ref{sec_strat}, the PW total static
structure factor does not recover, as $q\to 0$, the exact 
plasma frequency in its leading $q^2/2\omega_p$ term. 
This is visible for $r_s=8$ and 10.
\section{Correlation energy}
\label{sec_correnergy}
\subsection{Spin--unresolved}
The correlation energy obtained by integrating our
$g$ [see Eq.~(\ref{eq_excfromg})] 
is reported in Fig~\ref{fig_ec}, together with the
corresponding QMC data.\cite{OBH}
Its $\uparrow\uparrow$ and $\uparrow\downarrow$ contributions ($\epsilon_c=
\epsilon_c^{\uparrow\downarrow}+\epsilon_c^{\uparrow\uparrow}$)
are also separately shown. As expected,\cite{histo} correlations are
dominated by $\uparrow\downarrow$ interactions. Our total correlation
energies are in agreement with QMC data within 5\% (the maximum
absolute error is 3.4~mRy). Notice that, even if our model pair--distribution
function breaks down for $r_s>10$, it gives very good correlation
energies even at higher $r_s$ values. This is due to the optimal choice of
the $r_s$ dependence of our free parameters, which also includes the
low--density expansion of $\epsilon_c$. 

To have an idea of the accuracy of our correlation energies, we
performed best fits of the QMC data of  Ortiz, Harris and Ballone\cite{OBH}
(hereafter OHB)
based on other popular interpolation formulae for $\epsilon_c(r_s)$,
i. e. the Perdew--Zunger\cite{PZ}
(PZ), the Vosko--Wilk--Nusair\cite{Vosko} (VWN) and the
Perdew--Wang\cite{PWEc} (PW2, to distinguish it from
the pair--correlation model) functional forms.
The new QMC data for the correlation
energy of the unpolarized jellium are available for a large set of $r_s$:
0.8, 1, 2, 3, 4, 5, 8, 10, 20, 30, 40, 50 and 60. The results are the
followings: with the PZ formula one obtains a rather good fit (within 3\%),
but a wrong negative
coefficient for the high--density term $r_s\ln r_s$, an
unpleasant feature already pointed out in Ref.~\CITE{OB}. Moreover, the PZ
energy has a discontinuity in its second derivative at $r_s=1$,
an unpleasant feature for whoever is interested in the corresponding
pair--correlation function, related to the first derivative of
$\epsilon_c$.

The VWN form efficiently
interpolates the OHB data (2.7\% maximum relative error; 1.5~mRy maximum
absolute error) only if the free parameter $x_0$ of the VWN formula
has a positive value, which however implies an unphysical
logarithmic divergence at finite $r_s$ ($\sim 0.6$).
If $x_0$ is constrained to be negative,
then the fit provided by the VWN form is
not better than ours
(5.2\% maximum relative error; 3.4~mRy maximum absolute error).

The fit accomplished with the PW2
form is not very accurate (see also Ref.~\CITE{OB}):
7\% maximum relative error, 3.4~mRy maximum absolute error.
Moreover, the optimal fit parameter $\beta_4$ of PW2 form turns out to be
negative (see also Ref.~\CITE{OB}), thus leading to a negative coefficient
for the low--density expansion term $r_s^{-3/2}$ and to the
violation of the Ferrel condition.\cite{Ferrel}

We can conclude that the correlation energies vs. $r_s$,
which directly emerge from our pair--correlation
functions, although
not targeted by our fits, are rather good.

The main inaccuracies of the popular correlation--energy models just 
reviewed are located in the high--density region, where at first sight 
the new QMC results\cite{OBH,OB} cannot be reconciled with the exact 
$r_s\to 0$ limiting behavior.  
This discrepancy can be related to the combined impact of fixed--node 
approximation\cite{ballone} and infinite--size extrapolation (which 
would match the finding that, for $r_s\le 2$, Monte Carlo simulations 
based on different nodes and size--scaling rules Refs.~\CITE{cepald} 
and~\CITE{kwon} obtain somewhat different energies); it should be 
kept in mind, however, that the exact high--density expansion only 
holds for $r_s\to 0$, and could, in principle, start dominating the 
correlation energy at smaller $r_s$ values than implicitly 
assumed by the existing models.

An alternative correlation--energy model, not related to our 
pair--correlation function but capable of an excellent interpolation 
of the QMC energies of Refs.~\CITE{OB} and~\CITE{OBH} including
those at
high density, can be obtained by a minor generalization of the 
PW2 form.  Such a generalization keeps its exact 
$r_s\to 0$ limit, improves some of its original analytic 
properties, and appears flexible enough to interpolate different sets 
of high--density QMC data.\cite{cepald,OB,kwon,OBH} We separately 
present it in our Appendix~\ref{app_fitec}.
\subsection{Spin--resolved}
The spin-resolved contributions to the correlation energy,
shown in Fig.~\ref{fig_ec}, should be
reliable in the density range $r_s\le 10$, since they are
obtained by integrating the corresponding QMC pair--correlation functions.
This appears to be the only way to extract the $\uparrow\downarrow$ and
$\uparrow\uparrow$ contributions to $\epsilon_c$ from QMC data. 
For $r_s > 10$ we cannot expect our spin--resolved contributions to be
as reliable as for $r_s\le 10$, since at these very
low densities they do not correspond
to good pair--correlation functions (see Sec.~\ref{sec_results}). 

In Fig.~\ref{fig_ecuu} we compare
our parallel--spin part of the correlation energy with two corresponding 
widely--used scaling guesses: Perdew--Wang\cite{PerdewWang} 
[$\epsilon_c^{\uparrow\uparrow}(r_s,\zeta=0)=\epsilon_c(r_s,\zeta=1)/2^{1/3}$, where
$\zeta=|n_{\uparrow}-n{\downarrow}|/n$] and 
Stoll~{\it et~al.}\cite{Stoll} [$\epsilon_c^{\uparrow\uparrow}
(r_s,\zeta=0)=
\epsilon_c(2^{1/3}r_s,\zeta=1)$]. Both seem to overestimate the 
$\uparrow\uparrow$ contribution to the correlation energy.
Even if the Stoll~{\it et~al.}\cite{Stoll} estimate fulfills the exact
high--density limit\cite{WangPerdewEcHD} ($A/2\ln r_s$),  the 
PW\cite{PerdewWang} model (in which the $r_s\to 0$ limit is violated) 
seems to do better in the relevant density range $r_s\gtrsim 0.1$.

As $r_s$ increases, the PW and Stoll~{\it et~al.} approximations tend to 
the same limit, which is rather different from our result.
Fig.~\ref{fig_ecuu} suggests that,
even if we take a conservative approach and fully trust only our
$r_s \le 10$ spin--resolved contributions to $\epsilon_c$,  the
common PW and Stoll~{\it et~al.} low--density tail hardly matches
the QMC data.

\section{Conclusions and perspectives}
\label{concper}
We have proposed a new, analytic, spin--resolved, static structure
factor and
pair--correlation function for the unpolarized jellium
which works in the
density range $r_s\le 10$. Our model functions
fulfill a wealth of known analytic properties of their
exact counterparts, nicely interpolate the most recent
and complete QMC data of Ortiz, Harris and Ballone,\cite{OBH}
and consistently yield accurate correlation energies. They can be
of interest
to build up beyond--LDA exchange--correlation energy density
functionals,\cite{gunlun,GGA,WDA,Chacon,Dobson} for the magnetic response
of the unpolarized homogeneous electron gas,\cite{review,STLS_mag}
and also, within the theory developed in Refs.~\CITE{expO},
for the $e-e$ correlation in real materials. As a byproduct, we have
obtained two correlation energy models which work well in the entire
$r_s< \infty$ density range.

In further developments we plan to extend our procedure to the partially
polarized jellium and to lower densities ($r_s>10$).
A small Fortran code aimed at the numerical evaluation
of our functions [Eqs.~(\ref{Sud}), (\ref{Suu}),
(\ref{gud}), (\ref{guu}), (\ref{pippaPW})] can be obtained
upon request to {\tt Giovanni.Bachelet@roma1.infn.it}.

\section*{Acknowledgments}
We are very grateful to P. Ballone for making available to us prior 
to publication the numerical results of Ref.~\CITE{OBH}, on which 
this work is based, and to him and to J.~P. Perdew for a critical 
reading of the manuscript and many useful comments. We also thank 
D.~M. Ceperley, S. Conti, S. Moroni and G. Senatore for fruitful 
discussions. GBB gratefully acknowledges partial support from the
Italian Ministry for University, Research and Technology (MURST 
grant no.  9702265437).

\end{multicols}

\appendix
\section{Pair--correlation functions in real space}
\label{app_g}
The expressions of Eqs.~(\ref{Sud}) and~(\ref{Suu}) correspond in real
space to:
\begin{eqnarray}
g^c_{\uparrow\downarrow}(\rho;r_s) & = & \frac{\pi e^{-
a^{\uparrow\downarrow}\rho}}{480}
\biggl\{\frac{\alpha_4^{\uparrow\downarrow}}
{a^{\uparrow\downarrow}}\biggl[\frac{10395}{64}-\frac{12645}{64}
a^{\uparrow\downarrow}\rho+
 \frac{585}{8}(a^{\uparrow\downarrow}\rho)^2
-\frac{705}{64}(a^{\uparrow\downarrow}\rho)^3+\frac{45}{64}
(a^{\uparrow\downarrow}\rho)^4-\frac{(a^{\uparrow\downarrow}\rho)^5}
{64} \biggr]+ \nonumber \\
& & +\frac{\alpha_6^{\uparrow\downarrow}}{(a^{\uparrow\downarrow})^3}
\biggl[\frac{945}{64}+\frac{945}{64}a^{\uparrow\downarrow}\rho-
\frac{315}{16}(a^{\uparrow\downarrow}\rho)^2+\frac{345}{64}
(a^{\uparrow\downarrow}\rho)^3-\frac{33}{64}(a^{\uparrow\downarrow}\rho)^4+
\frac{(a^{\uparrow\downarrow}\rho)^5}{64}\biggr]\biggr\}+\nonumber \\
& & +3 \sum_{n=1}^6c^{\uparrow \downarrow}_n(-1)^{n+1}
\frac{\partial^{n+1}}{\partial
(b^{\uparrow \downarrow})^{n+1}} \left[ \frac{1}{\rho^2+
(b^{\uparrow \downarrow})^2} \right]
\label{gud} \\
g^c_{\uparrow\uparrow}(\rho;r_s) & = & \frac{\pi e^{-a^{\uparrow
\uparrow}\rho}}{6881280}\biggl\{\frac{\alpha_6^{\uparrow\uparrow}}
{(a^{\uparrow\uparrow})^3}\Big[135135+135135a^{\uparrow\uparrow}\rho-
270270(a^{\uparrow\uparrow}\rho)^2+114765(a^{\uparrow\uparrow}\rho)^3-
20370(a^{\uparrow\uparrow}\rho)^4+ \nonumber \\
& & 1722(a^{\uparrow\uparrow}\rho)^5
-68(a^{\uparrow\uparrow}\rho)^6+(a^{\uparrow\uparrow}\rho)^7\Big]+
\frac{\alpha_8^{\uparrow\uparrow}}{(a^{\uparrow\uparrow})^5}
\Big[-31185-31185a^{\uparrow\uparrow}\rho+6930(a^{\uparrow\uparrow}\rho)^2
+17325(a^{\uparrow\uparrow}\rho)^3+\nonumber \\
& & -6930(a^{\uparrow\uparrow}\rho)^4+
938(a^{\uparrow\uparrow}\rho)^5-52(a^{\uparrow\uparrow}\rho)^6
+(a^{\uparrow\uparrow}\rho)^7\Big]+\frac{\alpha_{10}^{\uparrow
\uparrow}}{(a^{\uparrow\uparrow})^7}\Big[14175+
14175a^{\uparrow\uparrow}\rho+1890(a^{\uparrow\uparrow}\rho)^2+
\nonumber \\
& & -2835(a^{\uparrow\uparrow}\rho)^3-882(a^{\uparrow\uparrow}\rho)^4+
378(a^{\uparrow\uparrow}\rho)^5-36(a^{\uparrow\uparrow}\rho)^6+
(a^{\uparrow\uparrow}\rho)^7\Big]\biggr\}+\nonumber \\
& & +3 \sum_{n=1}^6c^{\uparrow \uparrow}_n(-1)^{n+1}
\frac{\partial^{n+1}}{\partial
(b^{\uparrow \uparrow})^{n+1}} \left[ \frac{1}{\rho^2+
(b^{\uparrow \uparrow})^2} \right]
\label{guu}
\end{eqnarray}
\section{Optimal fit to the QMC correlation energy}
\label{app_fitec}
Since the Perdew--Wang\cite{PWEc} (PW2) form is 
simple and physically motivated, we slightly
modify it by introducing
one more free parameter which grants us enough flexibility to
accurately fit the new data by Ortiz, Harris and 
Ballone.\cite{OBH}

We also include the exact $r_s\ln r_s$ and $r_s$ coefficients
(see Subsec.~\ref{sub_EHD}) in 
the high--density expansion
of the functional form, that now reads:
\begin{equation}
\epsilon_c(r_s) = -2\,A\,(1+\alpha_1\,r_s+\alpha_2\,r_s^2)\,
 \ln\left(1+\frac{1}{2\,A\,\sum_{n=1}^6\beta_n\,r_s^{n/2}}\right).
\label{pippaPW}
\end{equation}
This modified PW2 form provides a much more drastic separation between the
high-- and low--density regime with respect to the original PW2 one. Such
a separation is crucial to obtain a good fit which both reproduces
the new QMC energies\cite{OBH} at the highest densities and avoids
undesired effects on the low--density regime (such as a negative
coefficient for the $r_s^{-3/2}$ term).
The parameters $A$, $\beta_1$, $\beta_2$, $\beta_3$ and $\alpha_1$ are fixed by
imposing the high--density expansion of Eq.~(\ref{Ec_HD}):
$\alpha_1=C/A$, $\beta_1=0.5/A \exp(0.5B/A)$,
$\beta_2=2A\beta_1^2$ and $\beta_3=0.5\beta_1(8\beta_1^2A^4-CB+DA)/A^2$. A best fit
to new QMC data\cite{OBH} gives for the 4 free parameters:
$\alpha_2=5$, $\beta_4=45$, $\beta_5=
32$, $\beta_6=12.7$. The resulting low--density expansion
is $-0.39/r_s+0.99/r_s^{3/2}$. The maximum absolute error is
1.6~mRy, while the maximum relative error is 2.4\%.

\begin{multicols}{2}

\end{multicols}

\vbox{
\begin{figure}
\begin{center}
\epsfxsize=11 truecm
\centerline{\epsfbox{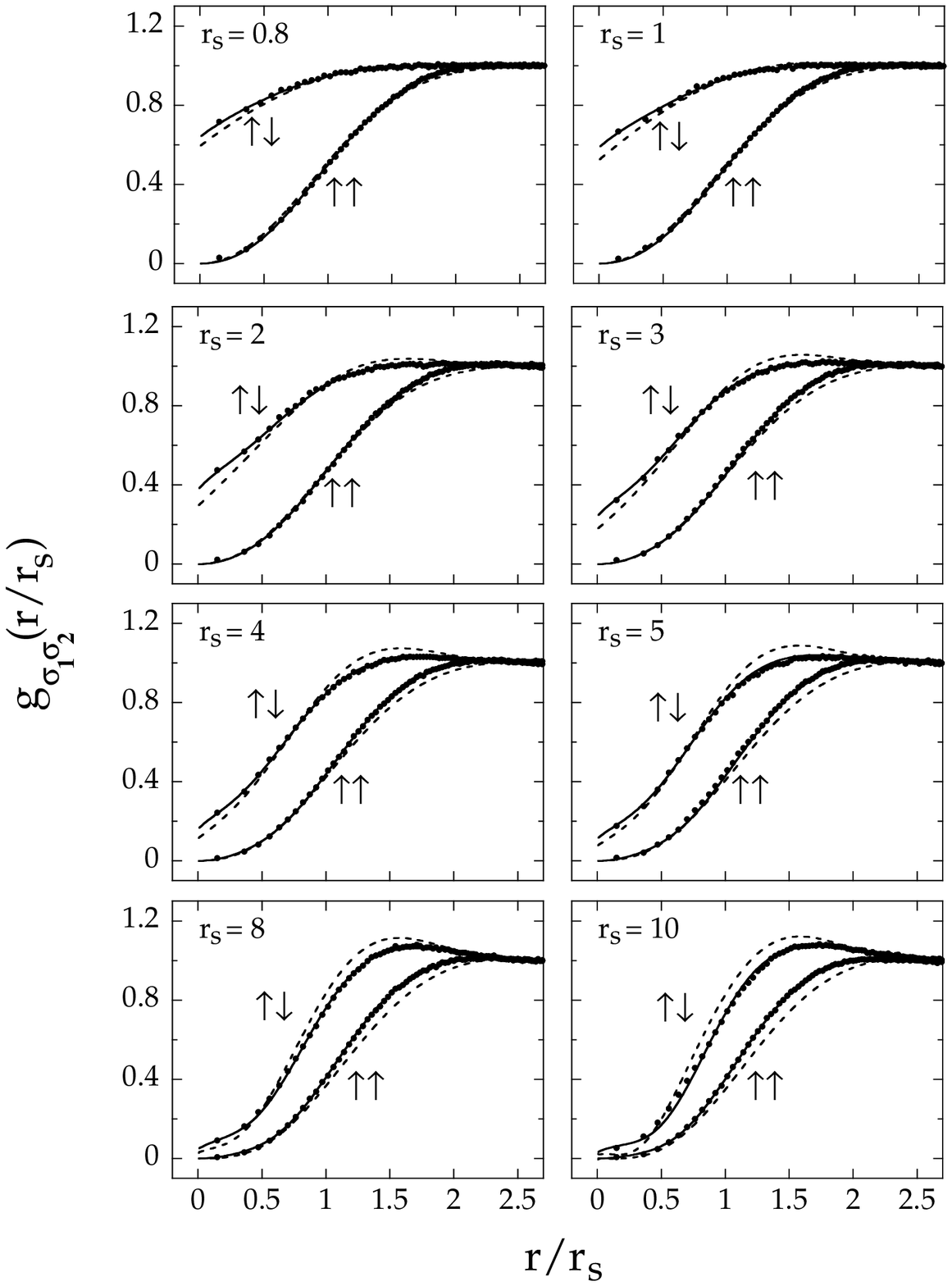}}
\end{center}
\caption{Spin--resolved pair--correlation function of the unpolarized
homogenous electron gas plotted against the electron separation $r$
scaled by the density parameter $r_s$ for eight different values of
$r_s$ between $r_s=0$ and $r_s=10$. Solid line: this work; dots: QMC data;
dashed line: Perdew--Wang model.}

\label{fig_g}
\end{figure}
}

\vbox{
\begin{figure}
\begin{center}
\epsfxsize=11 truecm
\centerline{\epsfbox{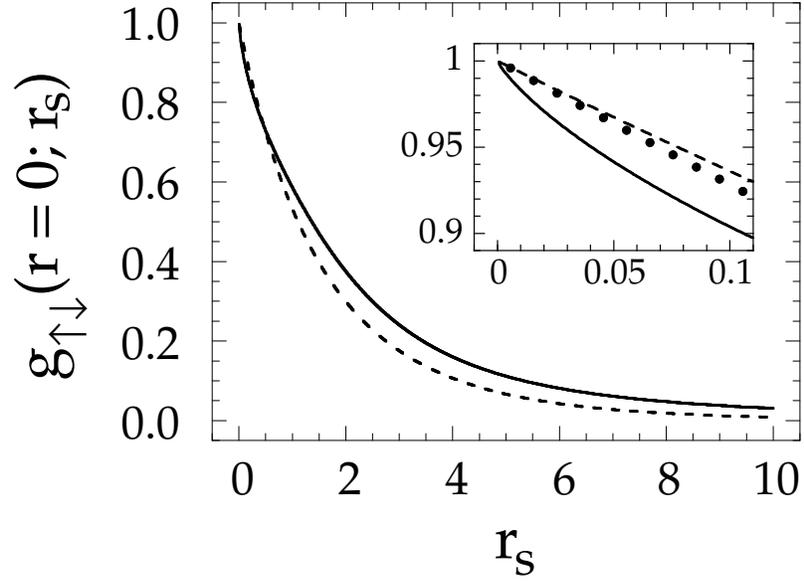}}
\end{center}
\caption{Antiparallel--spin correlation function at zero
interelectronic distance plotted against the density parameter $r_s$.
Solid line: this work; dashed line: electron--electron ladder
evaluation.\protect\cite{Yas} In the inset the exact high--density
expansion\protect\cite{KimballHD} is also shown as dots.}

\label{fig_g0}
\end{figure}
}

\vbox{
\begin{figure}
\begin{center}
\epsfxsize=11 truecm
\centerline{\epsfbox{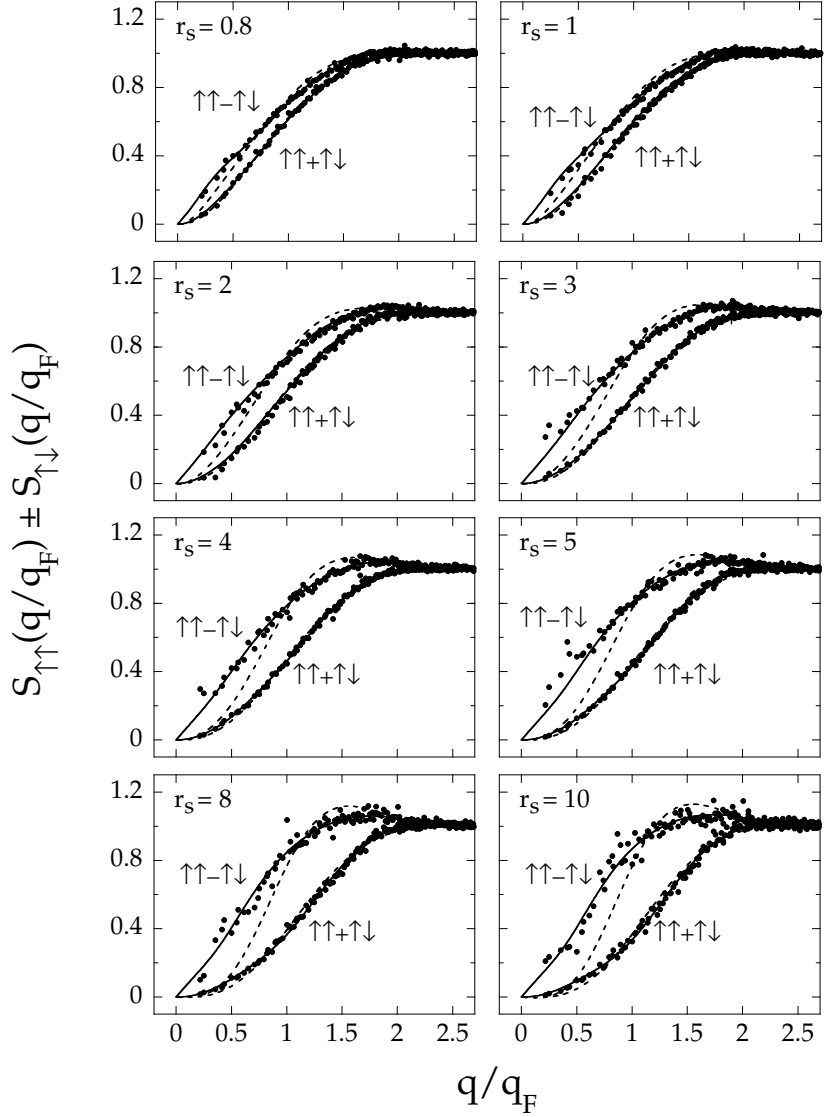}}
\end{center}
\caption{Static structure factor and magnetic
structure factor of the unpolarized
jellium plotted against the wavevector $q$
scaled by the Fermi wavevector $q_F$ for the same eight $r_s$
values as Fig.~\protect\ref{fig_g}. Solid line: this work; dots:
QMC data; dashed line: Perdew--Wang model.}

\label{fig_S}
\end{figure}
}

\vbox{
\begin{figure}
\begin{center}
\epsfxsize=11 truecm
\centerline{\epsfbox{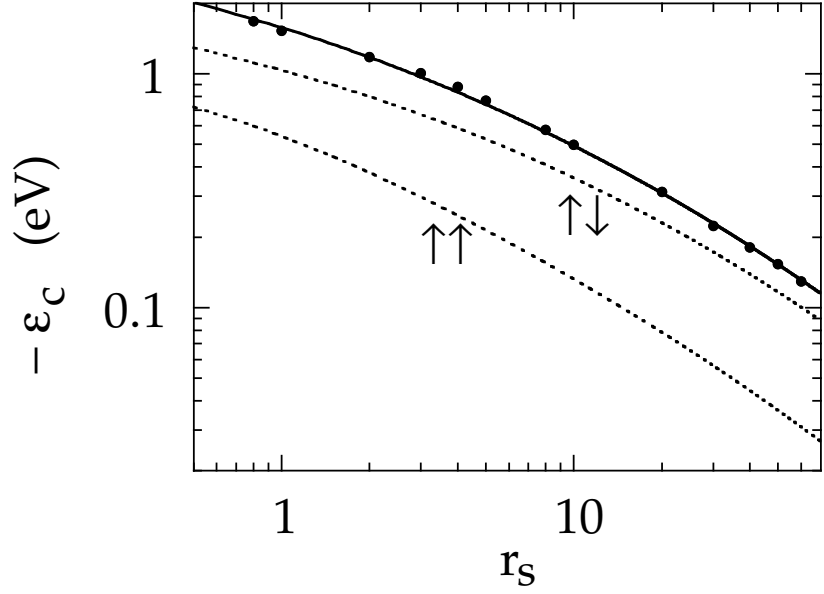}}
\end{center}
\caption{Total (solid) and spin--resolved (dashed) correlation energy of the
unpolarized homogenous electron gas plotted against the density
parameter $r_s$. QMC data are also shown as dots.}

\label{fig_ec}
\end{figure}
}

\vbox{
\begin{figure}
\begin{center}
\epsfxsize=11 truecm
\centerline{\epsfbox{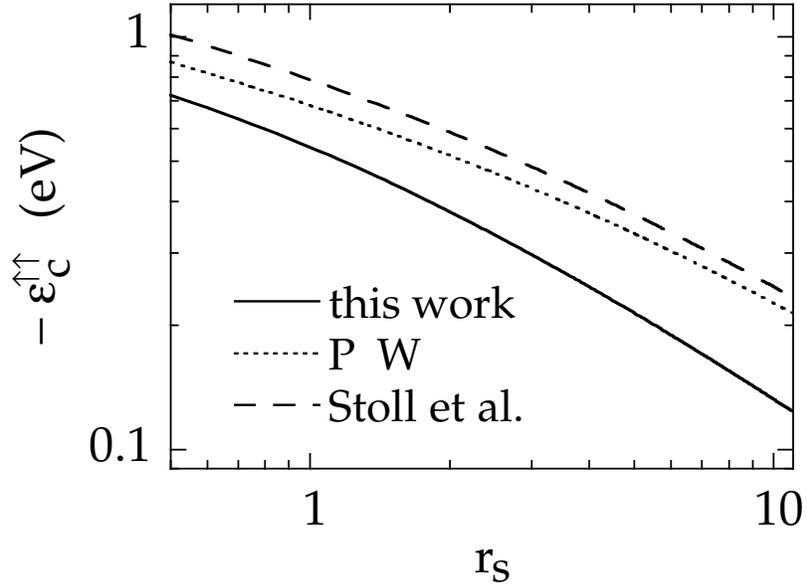}}
\end{center}
\caption{Our parallel-spin contribution to the total correlation energy 
compared to the Perdew--Wang\protect\cite{PerdewWang} and Stoll et 
al.\protect\cite{Stoll} approximations. $\epsilon_c(r_s,\zeta=1)$ is
from Ref.~\protect\CITE{PWEc}. }

\label{fig_ecuu}
\end{figure}
}


\begin{thebibliography}{100}
\bibitem{serieallucinante} the gaussian cut--off used by several
        authors~\protect\cite{OB,Kimball2,Ichimaru,Sacchetti,PerdewWang}
       in real or reciprocal space would,
        instead, give rise to unpleasant non--closed form
        espressions for the spherical Fourier transform of
        odd powers of $g(r)$ [and/or $S(q)$], since
        $\int_0^\infty e^{-ar^2} \sin (q r)dr=
        \frac{q}{2a}\sum_{n=1}^\infty\frac{1}{(2n-1)!!}\left(\frac{-q^2}{2a}
        \right)^{n-1}$.
\bibitem{newPW} we hear from J. P. Perdew that he and his collaborators are
       presently working on a better estimate for the spin resolution
       of the PW\protect\cite{PerdewWang} pair--correlation function.
\end{thebibliography}
\end{document}